# Mesoscale Elucidation of Biofilm Shear Behavior

Pallab Barai[1], Aloke Kumar[2] and Partha P. Mukherjee[1]

*Correspondence*: pmukherjee@tamu.edu (Partha. P. Mukherjee);

aloke.kumar@ualberta.ca (Aloke Kumar)

[1]Department of Mechanical Engineering, Texas A&M University, College Station, TX 77843, USA

[2]Department of Mechanical Engineering, University of Alberta, AB, Canada

# Abstract

Formation of bacterial colonies as biofilm on the surface/interface of various objects has the potential to impact not only human health and disease but also energy and environmental considerations. Biofilms can be regarded as soft materials, and comprehension of their shear response to external forces is a key element to the fundamental understanding. A mesoscale model has been presented in this article based on digitization of a biofilm microstructure. Its response under externally applied shear load is analyzed. Strain stiffening type behavior is readily observed under high strain loads due to the unfolding of chains within soft polymeric substrate. Sustained shear loading of the biofilm network results in strain localization along the diagonal direction. Rupture of the soft polymeric matrix can potentially reduce the intercellular interaction between the bacterial cells. Evolution of stiffness within the biofilm network under shear reveals two regions: a) initial increase in stiffness due to strain stiffening of polymer matrix, and b) eventual reduction in stiffness because of tear in polymeric substrate.

## Introduction

Bacteria can exist either as planktonic solutions or as surface associated colonies known as biofilms [1,2]. The latter is now recognized to be dominant mode of bacterial life. Biofilms generally include several differentiated populations of bacterial cells that are embedded in a matrix of self-produced extracellular polymeric substances (EPS). EPS acts as natural 'glue' providing mechanical integrity to the biofilm structure [3]. Due to this social form of growth biofilm bacteria offers resilience to external stresses and thus provides itself with an existential advantage over planktonic form of living [4-7]. Biofilms have been implicated for their role in recurrent infections and biofouling [8,9]. However, they also play a crucial role in several environmental processes including chemical cycles and waste-water treatment [10,11]. Our need to control biofilm growth, as well as possess the ability to remove undesirable biofilms, necessitates a proper understanding of the material characteristics of this biological soft matter [12].

Biofilms are now recognized to belong to the class of viscoelastic materials; this behavior dictates deformation behavior of biofilms under shear forces [13]. In practical situations such behavior leads to clogging of biomedical devices [14], porous media [15,16] as well as controls our ability to remove undesirable biofilms [17,18]. Experimental investigation of the viscoelastic properties of biofilms have been performed using techniques such as capillary flow cells, rotating disk rheometry, holographic microrheology, microbead force spectroscopy and also by using deformable microfluidic devices [4,5,7,19,20]. This has allowed researchers to quantify the elastic shear modulus as well the relaxation time of the viscoelastic material. Interestingly, there exists a wide variability in experimental results even when performed for the same strain [5,21,22]. Moreover, experiments have revealed interesting phenomena such as strain stiffening, which are poorly understood in the context of biofilms. Experimental evidence aside, a full understanding of biofilm properties remains elusive [6]. Heterogeneity in localization of cells and variability in EPS composition make rheological modeling of biofilms challenging. Moreover, EPS contains large biological macromolecules whose unfolding behavior can introduce complex stress-strain relationships [23-25]. Recent modeling efforts have used network theory or finite element modeling to capture biofilm mechanical behavior. It is imperative that open-ended questions such as those mentioned above be answered to move towards our goal of understanding biofilm deformation.

## Results and Discussion

Using a unique lattice spring based digital biofilm model (DBM), strain stiffening behavior observed in biofilm networks have been captured. Unfolding and subsequent stiffening of the protein chains located inside EPS resulted in increment of stiffness within the biofilm network. Since the bacteria located within the biofilm network acts as a rigid body, with increasing bacteria loading, effective stiffness of the biofilm network increases. Different portions of the same biofilm can have various bacteria loading. Figure 1(b), 1(c) and 1(d) shows three different cases of 26%, 42% and 55% bacteria loading, respectively.

Under shear deformation, the stress-strain curve obtained for different bacterial loading has been shown in Fig. 1(e). With increasing proportion of rigid bacteria, stiffness of the system increases. Strain stiffening is observed under all the three bacterial loading conditions. The spring stiffness parameter has been assigned in such a way that the stress-strain curve for the 26% bacteria loading corresponds to the experimental result reported by Stoodley et al. (2002) [5]. It can be observed from Fig. 1(e) that for higher bacteria loading, the strain stiffening starts earlier. An important point to note is that in our model, we assume that bacteria can be approximated as rigid particles [6]. Hence, under the application of an external force, majority of the deformation occurs within the EPS matrix because of its smaller stiffness. For the sections of biofilm with high bacteria loading also corresponds to low EPS content. All the strain stiffening comes only from the EPS matrix. The polymeric substrate in biofilm microstructure which contains smaller proportion of EPS experiences more strain, and the stiffening initiates earlier. On the other hand, polymeric substrates in biofilm microstructures containing a higher proportion of EPS are subjected to less strain, leading to delayed initiation of strain stiffening phenomena.

To check whether the stiffness values estimated using DBM is correct or not, a bound based analysis has been conducted in Fig. 1(f). The Hashin-Shtrikman (H-S) upper and lower bounds for shear modulus were plotted along with stiffness of the biofilm for all the three different bacteria loading at 5% and 35% shear strain [26]. The stiffness at 5% strain corresponds to the lower bound, whereas; at 35% strain the shear stiffness lie perfectly inside the H-S bounds. This implies that the H-S bounds for shear stiffness are applicable in biofilm networks as well for the purpose of checking the validity of experimental results.

It is conjectured that strain stiffening occurs within biofilm networks because of the unfolding of protein chains. The question is whether the entire EPS matrix experiences uniformly distributed shear strain, or localization of shear strain occurs within the DBM network. The biofilm morphology taken into consideration consists of 42% bacterial loading with uniform distribution (also described in the methodology section). Figure 2(a) shows (in cyan) the location of those springs which has experienced unfolding. It has been observed that very small amount of unfolding occurs on the left and right side of the biofilm network. Majority of the strain got localized along the diagonal direction which led to unfolding of springs along the diagonal. This type of shear strain localization along the diagonal has also been observed in metals [27]. From this analysis, it is evident that unfolding of the protein chains is localized along the diagonal of the EPS matrix. Rupture behavior has not been simulated in this case.

Force contours observed inside the biofilm network have been plotted in Fig. 2(b) and 2(c) under externally applied strains of 5% and 35%, respectively. For 5% strain, small amount of force has been observed within the network, which has a maximum value of 0.3N. As the strain is increased to 35%, large amount of force is generated, which shows a maximum value of 4.5N. It is clear that seven fold increment in strain (5% to 35%) results in fifteen fold increase in the magnitude of force (0.3N to 4.5N). This nonlinearity is attributed to the strain stiffening behavior observed due to the unfolding of EPS protein chains. Since unfolding of sub-springs lead to stiffening of the springs, an unfolded spring produces larger force. Thus, the unfolded springs along the diagonal direction observed in Fig. 2(a) naturally leads to localization of force along the diagonal (Fig. 2(b) and (c)).

None of the simulations discussed till now had rupture of the springs incorporated within the DBM. As rupture of springs is added into the simulations, the EPS matrix may experience detachment from the bacteria. Rupture of the EPS matrix can affect the chemical communication within the bacteria, which has the potential to impact the lifespan of a biofilm. By incorporating rupture of the protein chains, estimation will be made about the conditions when tearing can appear inside the EPS matrix. To simulate various bacterial morphologies, we used experimental images, which were later digitized. We considered various scenarios, such as, bacteria can be uniformly distributed (as shown in Fig. 3(a) and 3(d)) or extremely clustered at the center (see Fig. 3(b) and 3(e)). A third case of corner clustering has also been considered here to analyze the propagation of tear within the EPS matrix from one cluster to the other (see Fig. 3(c) and 3(f)).

Generation and nucleation of microcracks within uniformly distributed, centrally clustered and corner clustered bacteria are shown in Fig. 3(g), 3(h) and 3(i), respectively.

Density of ruptured protein chains within the EPS matrix is relatively less for the uniform distribution (see Fig. 3(g)). As a result some portion of every bacterium is connected with the EPS matrix. On the other hand, the biofilm morphology with centrally clustered bacteria experiences larger amounts of rupture inside the EPS matrix adjacent to the bacteria (see Fig. 3(h)). Extreme amount of tear within EPS has the potential to completely detach some of the bacterium from the biofilm network, which can impact their chemical communication. Thus high clustering of bacteria can be beneficial for removal of biofilm through the application of mechanical force. The physical reason behind observation of enhanced rupture in high clustering, stems from the fact that regions with high bacterial loading also experience lower concentration of EPS. Majority of the externally applied strain is accommodated by the soft EPS matrix. Strain stiffening is observed earlier in regions with less amount of polymeric substrate, which subsequently leads to rupture. This phenomenon is similar to that observed in Fig. 1(e).

Due to the presence of larger volume fraction of EPS matrix adjacent to each bacterium, less strain stiffening and rupture occurs under uniform distribution. The corner clustering case has been investigated to understand the propagation of tear through the EPS matrix in between two clusters. It is evident from Fig. 3(i) that the cracks through the EPS matrix propagates from one cluster to the nearest neighboring cluster. Probability of tear propagation along the diagonal direction is relatively less. Rupture propagation is a mechanism to release strain energy. Connecting the ruptures in the nearest neighboring clusters provides a shorter pathway to release strain energy quickly. That is why no rupture patterns have been observed which connects diagonally opposite clusters.

Stress vs. strain response of the biofilm network under shear strain induced loading has been shown in Fig. 4(a). Mechanical rupture has also been incorporated within the simulation. Due to unfolding of springs shear stiffness increases initially. Once the rupture of EPS matrix is initiated, stiffness of the network decreases significantly. Stress-strain curve for both the central clustering and uniformly distributed bacteria loading has been shown. The result obtained from computational analysis corresponds very closely with that observed in experiments (see Fig. 4(a))[19]. Because of localized strain stiffening, early initiation of rupture has been observed in the central cluster. For this particular case, cracks initiate at a stress level of $1N/m^2$. For uniformly distributed bacteria loading, rupture initiation starts at a stress level of $1.2N/m^2$. Delay in microcrack nucleation happens because of the fact that uniformly distributed bacteria contains large amount of EPS matrix adjacent to itself which acts as a buffer against rupture initiation.

Evolution of shear stiffness has been plotted in Fig. 4(b) with respect to increasing shear strain. It has been observed that strain stiffening dominates till a particular strain. Because of the presence of rupture in the protein chains of EPS matrix, rupture behavior starts to dominate after that critical strain. Two regions where strain stiffening dominates, and where rupture dominates have been clearly marked in Fig. 4(b). For the case of central clustering, domination of rupture over strain stiffening starts earlier than the case of uniformly distributed bacteria. This early initiation of rupture in the EPS matrix is attributed to the fact that in centrally clustered biofilm very little amount of EPS matrix exist adjacent to the bacterium. Under shear induced deformation, the EPS matrix carries majority of the strain and experiences early initiation of strain stiffening and subsequent rupture in the regions with less amount of the polymeric substrate. As a result, the uniformly distributed bacteria experiences delayed initiation of strain stiffening and rupture due to the presence of excessive amount of EPS matrix around each bacterium.

**Conclusion**

In this article, a novel digital biofilm model has been developed, which has the potential to take into consideration the effect of any bacteria–EPS microstructure observed in real biofilms. Different volume fractions of bacterial cells show that for higher bacteria loading, stiffer stress-strain response is obtained. Also, under high bacteria loading enhanced strain hardening occurs within EPS. It leads to earlier initiation of tears inside the protein chains. Similar to strain localization observed in metals, localized strain stiffening along the diagonal direction has been observed in biofilm networks under shear deformation. Enhanced rupture has been observed in EPS matrix which is adjacent to the bacteria in a closely packed cluster. As a result, detachment of the bacterial cells from the EPS matrix is possible in a highly clustered domain under mechanically induced shear type load. By plotting the evolution of stiffness in a biofilm network, two different regions have been recognized. In the first domain strain stiffening dominates and the effective stiffness of the biofilm network increases. In the second region, the effect of rupture of the polymer matrix plays a major role and eventually results in reduction of the effective stiffness.

**Methodology**

In this article, morphology aware computational model has been developed that can elucidate the shear behavior of a biofilm. In order to investigate the effect of shear force on a biofilm (Fig 1a), we begin by using experimental visualization of the biofilm structure to create a digital representation of the biofilm (Fig. 1(b-d)). Once the digital representation of the biofilm is complete, then a single domain modeling approach is applied by superimposing the digitized biofilm microstructure on top of a network of lattice springs (see Fig. 5(b)). This computational characterization of a realistic biofilm microstructure has been named as Digital Biofilm Model (DBM).

As shown in Fig. 5(a), in the digital biofilm model a triangular network of lattice springs has been considered to analyze the mechanical deformation of the biofilm. In the quasistatic computational scheme adopted in the current research, entire mass of the system is assumed to be concentrated at uniformly distributed nodes [28]. Any two adjacent nodes are connected by a spring, which shows only axial resistance. The triangular spring network leads to a coordination number of six. Interaction with only the nearest neighbors has been taken into consideration. Because of the triangular configuration, Poisson's ratio of 1/3 has been observed for this network.

Superimposing the digitized biofilm on top of the triangular network, springs which lie within the bacteria and springs which reside inside the EPS matrix has been separated (see Fig. 5(b)). In order to estimate stiffness of the biofilm network, the bacteria were assumed to be rigid with respect to the extracellular polymeric substrate (EPS) [6]. Furthermore, we restrict ourselves to studying biofilm response to shear forces for time scales much smaller than the viscoelastic relaxation time scales (i.e. $t \ll t_{relaxation}$) [5]. The elastic behavior dominates biofilm response in such small time scale. The EPS matrix consists of protein chains, which show strain-stiffening behavior due to unfolding of the polymer-chains [24]. Since the biofilm consists of soft polymeric gels, due to external effects the film can go through large deformations. Sometime shear strains as large as 50% can be observed because of flow induced force [7].

The rules applied to the spring are: (a) springs which are representative of bacteria have a very high spring constant [6], (b) springs which are in the EPS have a low spring constant and an energy prescribed limit is imposed. When the threshold strain energy is exceeded the springs become 'uncoiled' and are represented by a higher magnitude of spring constant. This latter rule is inspired by findings of Buehler [23,24]. Using molecular dynamic simulations, unfolding of the helical patterns were captured which leads to stiffening of the entire polymer chain. The untangling mechanism was initiated by the stutter defect present within a protein chain. Once all the protein chains go through the unfolding mechanism, the extended fiber show relatively high stiffness, but cannot sustain a significant rise in externally applied strain. (c) Once the strain energy stored in a spring exceeds its rupture threshold, the spring breaks. It is irreversibly removed from the network of lattice springs and the force carried by this spring gets distributed among its neighbors [29].

To capture the unfolding of springs, it is assumed that each spring in the network consists of eight folded sub-springs [30]. Due to application of shear strain, as each of the sub-springs unfold stiffness of the full spring increases (see Fig. 5(c)). After the eighth sub-spring has unfolded, application of more external load leads to rupture of the entire spring [31]. The remaining question is how to determine when the unfolding of sub-springs will occur. An energy prescribed criteria has been adopted in this study. Under externally applied loads, the local force $(f)$ and displacement $(u)$ has been calculated which equilibrate the spring network. Strain energy in each spring is calculated as, $\psi = (1/2) f \cdot u$. With increasing amount of externally applied load, the total strain energy stored in each spring increases. Eight uniformly distributed random energy thresholds have been generated for each spring $(\psi_{t,i}; i = 1,2,3,...,8)$ in such a way that $\psi_1 < \psi_2 <$

$\psi_3 < \ldots < \psi_8$. As the strain energy in the full spring exceeds an unfolding threshold $(\psi > \psi_{t,i})$, one sub-spring unfolds and the stiffness of the full spring is increased (shown in Fig. 5(c)). This untangling mechanism continues until all the eight sub-springs have unfolded. Stiffness of the spring reaches its maximum value at that point [24].

Another energy threshold has been assigned to all the springs which corresponds to its ultimate strength $(\psi_{t,u})$ and it follows the constraint $\psi_{t,u} > \psi_8$. Once the strain energy in the completely unfolded spring exceeds its ultimate rupture threshold $(\psi > \psi_{t,u})$, the spring is assumed to be broken. Hence, it is irreversibly removed from the lattice network and the contribution of the spring is completely eliminated from the stiffness matrix. Rupture in the lattice spring network results in reduction in stiffness of the entire biofilm. Since the spring constant of the full spring increases with each unfolding of the sub-springs, the stiffness of the lattice network also changes. In order to correctly capture the increase in stiffness and model the spring deformation accordingly, an iterative technique has been adopted. The externally applied load is increased in increments of very small magnitude. All the sub-springs that unfold or full springs that break within a loading interval are modified accordingly. The energy in each springs are also estimated in increments, $\psi^{n+1} = \psi^n + (1/2)\Delta \boldsymbol{f} \cdot \Delta \boldsymbol{u}$. Here, $\psi^{n+1}$ and $\psi^n$ signifies the strain energy at the current and the previous steps, respectively. $\Delta \boldsymbol{f}$ and $\Delta \boldsymbol{u}$ corresponds to the incremental force and displacement, respectively, observed in the current loading interval.

## Acknowledgements

Financial support from Texas A&M University faculty research initiation grant is gratefully acknowledged.

## Author contributions

A.K. and P.P.M. conceived the problem. P.B., A. K. and P.P.M. developed the theory. P.B. generated the result. P.B., A.K. and P.P.M. wrote the manuscript.

## Competing financial interests

The authors declare no competing financial interests.

## List of figures

**Figure 1.** This figure shows different amount of bacteria loading and the corresponding stress – strain response. (a) A representative biofilm microstructure subjected to shear load. (b) Bacteria loading of 26% (white portion signifies bacteria and black signifies the EPS matrix). (c) Bacteria loading of 42%. (d) Bacteria loading of 55%. (e) Shear stress – strain response of the biofilm for the three types of bacteria loading scenario. Since the bacteria are considered as rigid objects, larger volume fraction of bacteria leads to higher stiffness for the biofilm. Stiffness for each lattice spring element in the EPS matrix has been set in such a way that the biofilm with 26% bacteria loading follow the experimental shear stress – strain data obtained from Stoodley et al (2002). (f) Based on the stiffness of the EPS matrix and the bacterium, Hashin – Shtrikman (H-S) lower and upper bounds of shear stiffness for the biofilm has been evaluated. Shear stiffness of the biofilm at 5% and 35% shear strain falls within the Hashin – Shtrikman bounds.

**Figure 2.** Force contours depict the corresponding regions which experience strain hardening. No rupture has been considered in this simulation. (a) The cyan region signifies the portion which has undergone strain hardening behavior. Under shear loading, maximum shear strain has been observed along the diagonal directions. This is equivalent to the formation of shear bands observed in conventional solid materials. (b) Force contour observed at 5% shear strain. The magnitude of force is in the order of 0.3N. (c) Force contour observed at 35% shear strain. Magnitude of force is around 4.5N, which is an order of magnitude greater than the forces observed under 5% strain. Larger amount of forces accumulate along diagonal directions, which eventually leads to enhanced strain hardening along the diagonal direction. Similar behavior has been reported in part (b) of this image.

**Figure3.** Different clustering of bacteria inside the biofilm along with the rupture patterns are displayed here. (a) Uniform distribution of bacteria within a biofilm. (b) Clustering of bacteria in the central portion of the domain. (c) Clustering of bacteria is observed around the corner of the domains. (d) Schematic representation of uniformly distributed bacteria inside biofilms. (e)

Schematic representation of central clustering observed in biofilms. (f) Schematic representation of corner clustering within biofilms. (g) Distribution of rupture within the uniformly distributed bacteria. When all the bacteria are well dispersed, damage in the biofilm is not localized enough for an entire bacterium to be completely detached from the EPS matrix. (h) Rupture profile as observed for centrally clustered bacteria. When the bacteria are clustered within the biofilm, force concentration is very high around the bacterium which leads to mechanical detachment of the EPS matrix from certain bacterium. (i) Evolution of damage for corner clustering scenario. Purpose of the corner clustering is to investigate how the tear would propagate through the EPS between two clusters of bacteria. It can be concluded that cracks through EPS matrix propagates to the nearest cluster. There is very little possibility for a tear through EPS to span between two clusters which are located far away from each other.

**Figure 4.** Evolution of stress and stiffness for a biofilm under shear loading. (a) Shear stress is plotted against shear strain. For biofilms with bacterial clustering, rupture initiates at a stress level of approximately $1 N/m^2$. Whereas, for biofilms with uniformly dispersed bacteria, rupture initiates at a higher stress magnitude, around $1.2\ N/m^2$. (b) Evolution of shear stiffness of the biofilm under increasing shear strain. Initially, because of strain hardening, stiffness of the biofilm increases. But after sometime, rupture starts to dominate and the overall stiffness of the biofilm drops. The two regions are clearly demonstrated in the figure, where, strain hardening dominates initially and rupture takes over towards the end. Reduction in stiffness starts earlier for biofilms with central clustering. This means, it is easier to tear the biofilms with central clustering as compared to biofilms with uniformly distributed bacteria.

**Figure 5.** These figures demonstrate a schematic diagram of the lattice spring network, digitization of the biofilm domain and a schematic representation of the unfolding of springs. (a) The lattice spring network consists of a triangular mesh of spring elements. Each spring shows only axial stiffness. The mass remains concentrated at each node. The bottom of the network is fixed and shear force is applied at the top. (b) The digitized biofilm domain is displayed here. The red boundaries signify the boundaries for each of the bacteria. Blue portion is the mesh used to discretize the biofilm domain. (c) A schematic representation of the unfolding of springs. Under externally applied load, as the energy in a spring exceeds its unfolding threshold (the spring in the middle), it becomes straight and displays an infinite stiffness.

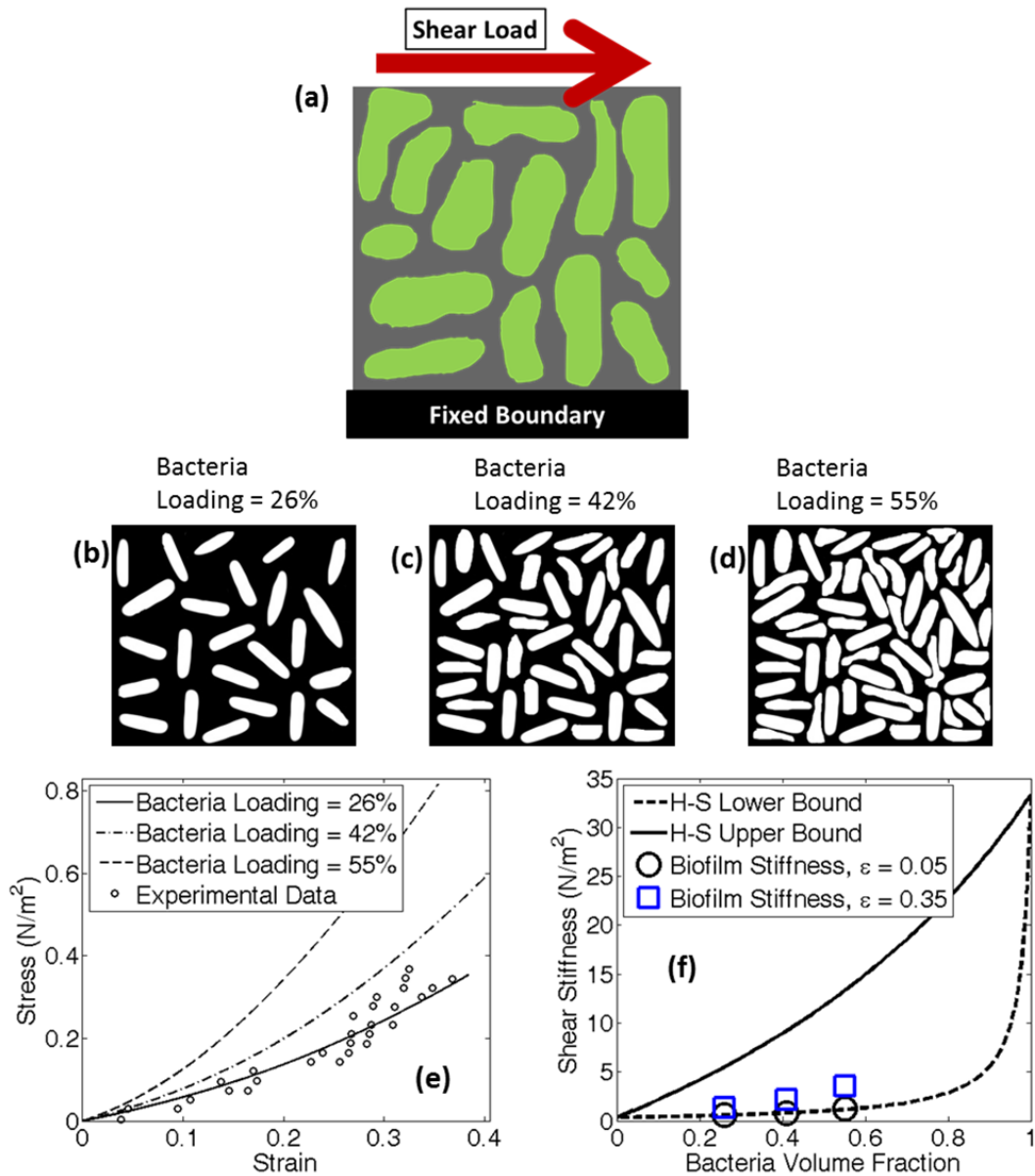

**Figure: 1.** This figure shows different amount of bacteria loading and the corresponding stress – strain response. (a) A representative biofilm microstructure subjected to shear load. (b) Bacteria loading of 26% (white portion signifies bacteria and black signifies the EPS matrix). (c) Bacteria loading of 42%. (d) Bacteria loading of 55%. (e) Shear stress – strain response of the biofilm for the three types of bacteria loading scenario. Since the bacteria are considered as rigid objects, larger volume fraction of bacteria leads to higher stiffness for the biofilm. Stiffness for each lattice spring element in the EPS matrix has been set in such a way that the biofilm with 26% bacteria loading follow the experimental shear stress – strain data obtained from Stoodley et al (2002). (f) Based on the stiffness of the EPS matrix and the bacterium, Hashin – Shtrikman (H-S) lower and upper bounds of shear stiffness for the biofilm has been evaluated. Shear stiffness of the biofilm at 5% and 35% shear strain falls within the Hashin – Shtrikman bounds.

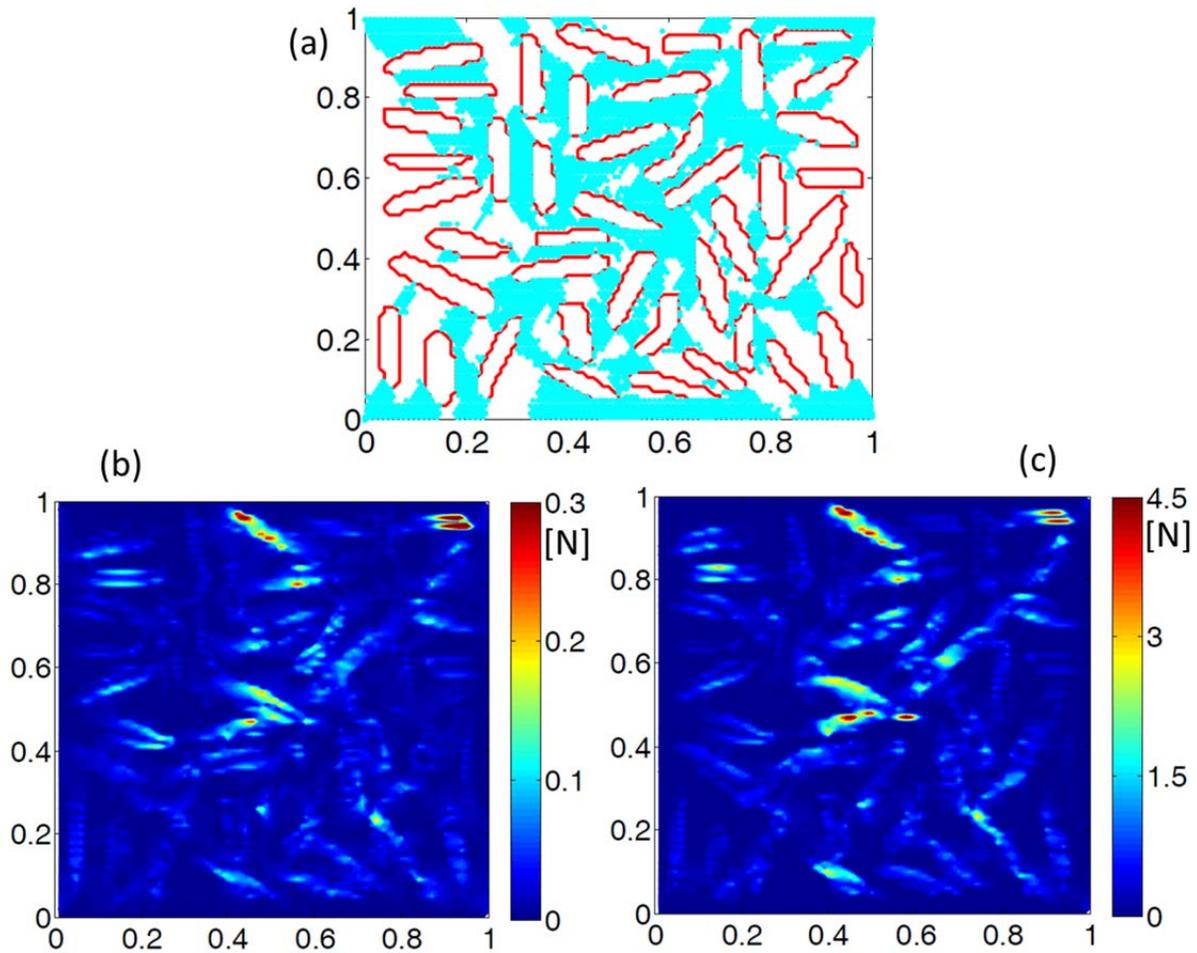

**Figure: 2.** Force contours depict the corresponding regions which experience strain hardening. No rupture has been considered in this simulation. (a) The cyan region signifies the portion which has undergone strain hardening behavior. Under shear loading, maximum shear strain has been observed along the diagonal directions. This is equivalent to the formation of shear bands observed in conventional solid materials. (b) Force contour observed at 5% shear strain. The magnitude of force is in the order of 0.3N. (c) Force contour observed at 35% shear strain. Magnitude of force is around 4.5N, which is an order of magnitude greater than the forces observed under 5% strain. Larger amount of forces accumulate along diagonal directions, which eventually leads to enhanced strain hardening along the diagonal direction. Similar behavior has been reported in part (b) of this image.

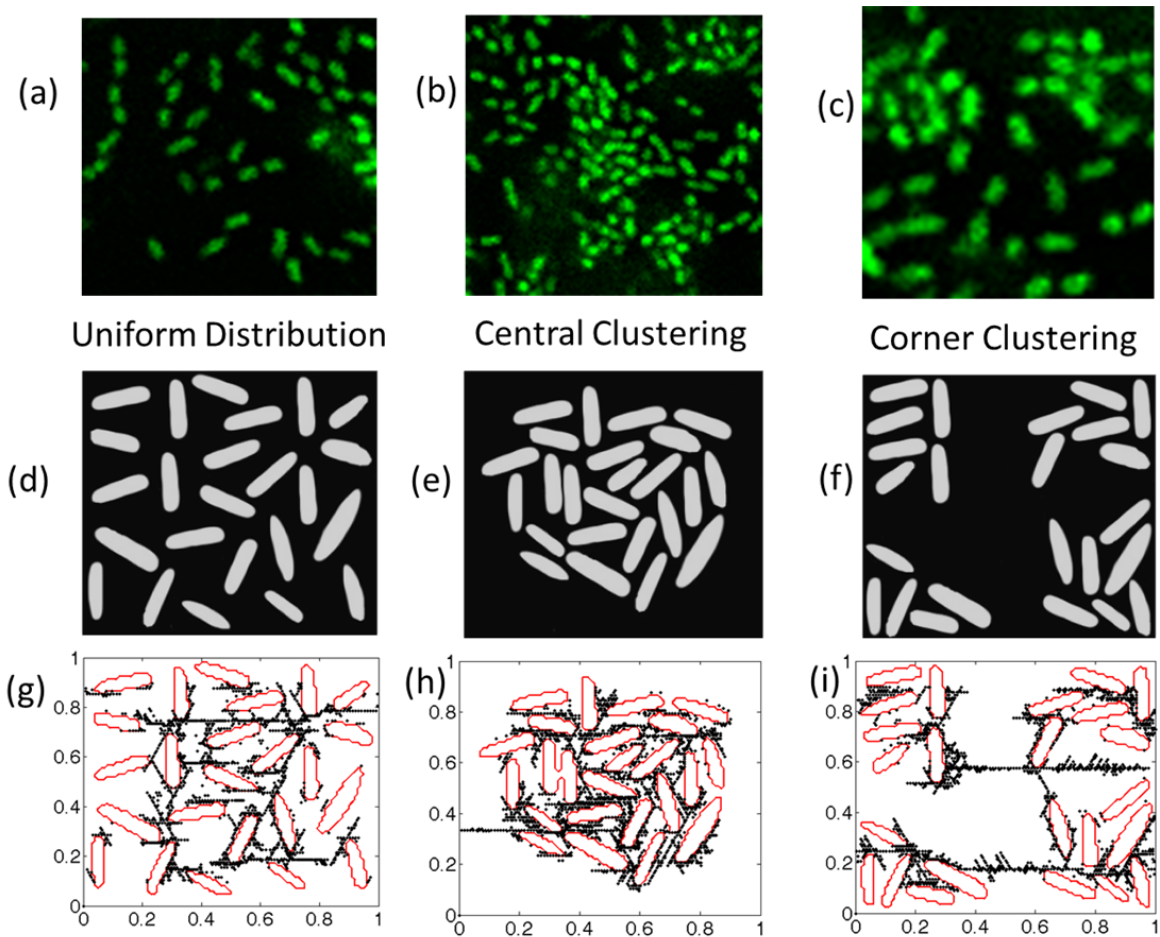

**Figure: 3.** Different clustering of bacteria inside the biofilm along with the rupture patterns are displayed here. (a) Uniform distribution of bacteria within a biofilm. (b) Clustering of bacteria in the central portion of the domain. (c) Clustering of bacteria is observed around the corner of the domains. (d) Schematic representation of uniformly distributed bacteria inside biofilms. (e) Schematic representation of central clustering observed in biofilms. (f) Schematic representation of corner clustering within biofilms. (g) Distribution of rupture within the uniformly distributed bacteria. When all the bacteria are well dispersed, damage in the biofilm is not localized enough for an entire bacterium to be completely detached from the EPS matrix. (h) Rupture profile as observed for centrally clustered bacteria. When the bacteria are clustered within the biofilm, force concentration is very high around the bacterium which leads to mechanical detachment of the EPS matrix from certain bacterium. (i) Evolution of damage for corner clustering scenario. Purpose of the corner clustering is to investigate how the tear would propagate through the EPS between two clusters of bacteria. It can be concluded that cracks through EPS matrix propagates to the nearest cluster. There is very little possibility for a tear through EPS to span between two clusters which are located far away from each other.

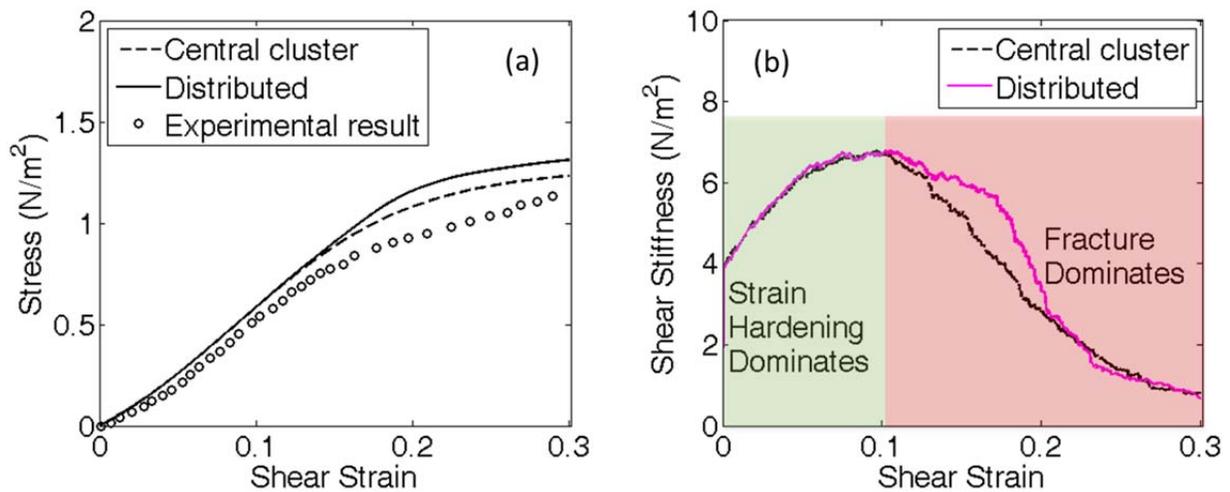

**Figure: 4.** Evolution of stress and stiffness for a biofilm under shear loading. (a) Shear stress is plotted against shear strain. For biofilms with bacterial clustering, rupture initiates at a stress level of approximately 1N/m$^2$. Whereas, for biofilms with uniformly dispersed bacteria, rupture initiates at a higher stress magnitude, around 1.2 N/m$^2$. (b) Evolution of shear stiffness of the biofilm under increasing shear strain. Initially, because of strain hardening, stiffness of the biofilm increases. But after sometime, rupture starts to dominate and the overall stiffness of the biofilm drops. The two regions are clearly demonstrated in the figure, where, strain hardening dominates initially and rupture takes over towards the end. Reduction in stiffness starts earlier for biofilms with central clustering. This means, it is easier to tear the biofilms with central clustering as compared to biofilms with uniformly distributed bacteria.

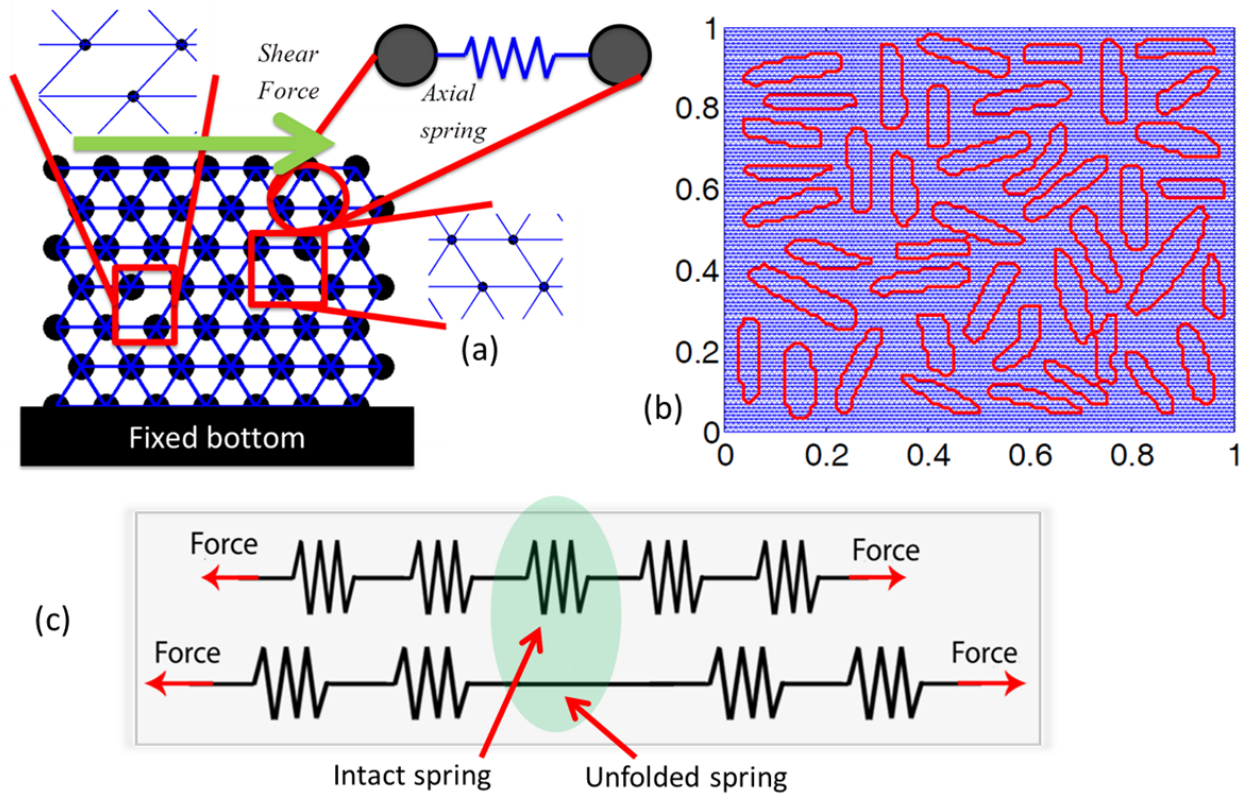

**Figure: 5.** These figures demonstrate a schematic diagram of the lattice spring network, digitization of the biofilm domain and a schematic representation of the unfolding of springs. (a) The lattice spring network consists of a triangular mesh of spring elements. Each spring shows only axial stiffness. The mass remains concentrated at each node. The bottom of the network is fixed and shear force is applied at the top. (b) The digitized biofilm domain is displayed here. The red boundaries signify the boundaries for each of the bacteria. Blue portion is the mesh used to discretize the biofilm domain. (c) A schematic representation of the unfolding of springs. Under externally applied load, as the energy in a spring exceeds its unfolding threshold (the spring in the middle), it becomes straight and displays an infinite stiffness.